# Collaborative error-immune information transfer with OAM modes of light in networks through atmospheric turbulence

*Rajni Bala\*, Sooryansh Asthana,  V. Ravishankar*

*Department of Physics, IIT Delhi, Hauz Khas, New Delhi, 110016, Delhi, India*

\**Rajni.Bala@physics.iitd.ac.in*

**Keywords:** crosstalk, error-immune information, turbulence, orbital angular momentum, quantum communication

**Abstract:** We present a protocol for collaborative error-immune information through atmospheric turbulence. The protocol allows to transfer information in a network of three participants.

## I. Introduction:

Recently, the field of quantum communication has witnessed an unprecedented research interest [1]. There have been various proposals for information transfer protocols. The experimental realisations of these protocols suffer greatly from channel noises [2]. Though there have already been many solutions in terms of quantum error-correcting and error-rejecting codes, they employ costly resources such as multiparty entanglement [3]. The generation of multipartite entangled states is yet in its infancy [4]. Keeping these in mind, in a recent work, we have proposed a new information encoding scheme for combating errors [5]. In this scheme, information is encoded in invariants which are special combinations of expectation values of observables. For a given noisy channel, by encoding information in these invariants, it can be transferred in an error-immune manner.

In recent times, orbital angular momentum states of light are at the forefront in experimental realisations of many communication protocols [6]. However, transmission of these states in free space suffers from atmospheric turbulence [7]. In a subsequent work, employing the framework proposed in [5], we have identified invariants for propagation of OAM modes of light passing through weak atmospheric turbulence [8]. In this work, using these invariants, we propose a protocol for collaborative information transfer in a network of three participants. The protocol allows to transfer information even after OAM modes of light pass-through atmospheric turbulence. The transferred information can be retrieved only if all the three participants collaborate with each other. Since the proposed protocol does not employ any additional costly resources, it is realisable with current technologies.

## II. Invariants for OAM modes in weak atmospheric turbulence:

In [7], the propagation of `three-qubit' OAM entangled state through atmospheric turbulence is simulated. The state chosen to be,

$$|\psi\rangle = \alpha_1|l,-l,-l\rangle + \alpha_2|-l,l,-l\rangle + \alpha_3|-l,-l,l\rangle + \alpha_4|l,l,l\rangle + \alpha_5|-l,-l,-l\rangle, \quad \cdots (1)$$

where $|\pm l\rangle$ represents a state with OAM mode index $\pm l$. In [7], it is assumed that the three photons suffer the same turbulence. Due to the effect of turbulence, the final state is a mixed state, represented by the density matrix ρ as given in [7]. The noisy parameters $(a, b)$ in the final state are survival and crosstalk amplitudes respectively.

To construct the invariants, which will be employed in error immune collaborative information transfer protocol, we, first define the set of observables,

$$\sigma_x^j = |l\rangle\langle -l| + |-l\rangle\langle l|, \sigma_y^j = i|l\rangle\langle -l| - i|-l\rangle\langle l|,$$

$$\sigma_z^j = |l\rangle\langle l| - |-l\rangle\langle -l|, \qquad j \epsilon (A, B, C)$$

Of particular interest to us are the following invariants for this channel [8]:

$$I_1 = \frac{\langle (\sigma_y^A \sigma_x^C - \sigma_x^A \sigma_y^C)(1^B - \sigma_z^B) \rangle}{\langle (\sigma_x^A \sigma_x^C + \sigma_x^A \sigma_x^C)(1^B - \sigma_z^B) \rangle}, I_2 = \frac{\langle (1^A + \sigma_z^A)(\sigma_x^B \sigma_y^C - \sigma_y^B \sigma_x^C) \rangle}{\langle (1^A - \sigma_z^A)(\sigma_x^B \sigma_x^C - \sigma_y^B \sigma_y^C) \rangle}$$

$$I_3 = \frac{\langle (\sigma_x^A \sigma_y^B - \sigma_y^A \sigma_x^B)(1^C - \sigma_z^C) \rangle}{\langle (\sigma_x^A \sigma_x^B + \sigma_x^A \sigma_x^B)(1^C - \sigma_z^C) \rangle}, \quad I_4 = \frac{\langle (\sigma_x^A)(1^B + \sigma_z^B)(1^C + \sigma_z^C) \rangle}{\langle (\sigma_y^A)(1^B + \sigma_z^B)(1^C + \sigma_z^C) \rangle}$$

$$I_5 = \frac{\langle (\sigma_x^A + i\sigma_y^A)(\sigma_x^B - i\sigma_y^B)(1^C - \sigma_z^C) \rangle}{\langle (\sigma_x^A + i\sigma_y^A)(1^B - \sigma_z^B)(\sigma_x^C - i\sigma_y^C) \rangle}, \quad I_6 = \frac{\langle (\sigma_x^A + i\sigma_y^A)(\sigma_x^B - i\sigma_y^B)(1^C - \sigma_z^C) \rangle}{\langle (1^A + \sigma_z^A)(\sigma_x^B - i\sigma_y^B)(\sigma_x^C - i\sigma_y^C) \rangle}$$

$$I_7 = \frac{\langle (\sigma_x^A + i\sigma_y^A)(\sigma_x^B - i\sigma_y^B)(1^C - \sigma_z^C) \rangle}{\langle (\sigma_x^A + i\sigma_y^A)(1^B - \sigma_z^B)(1^C - \sigma_z^C) \rangle}, \quad I_8 = \frac{\langle (\sigma_x^A + i\sigma_y^A)(1^B - \sigma_z^B)(\sigma_x^C - i\sigma_y^C) \rangle}{\langle (1^A - \sigma_z^A)(\sigma_x^B - i\sigma_y^B)(\sigma_x^C + i\sigma_y^C) \rangle}.$$

We now move on to propose an error-immune information transfer protocol with OAM modes of light passing through atmospheric turbulence.

### III. Collaborative error-immune information transfer in a network:

Suppose that David wish to transfer information in a network of three participants, *viz.*, Alice, Bob and Charlie. For this purpose, he employs a `three-qubit' OAM state $|\psi\rangle$ given in equation (1). Since each photon passes through atmospheric turbulence, David encodes information in the invariants. The steps of the protocol are as follows:

1. David prepares the state $|\psi\rangle$ and sends the first, the second, and the third subsystem to Alice, Bob, and Charlie.
2. Alice, Bob, and Charlie upon receiving the state randomly perform measurements of $\sigma_x^j, \sigma_y^j, \sigma_z^j$ on their respective subsystems. These measurements can be performed using spatial light modulators.
3. After a sufficient number of rounds, Alice, Bob, and Charlie communicate among themselves to retrieve information.
4. All the three participants communicate their measurement bases and corresponding results.

In this manner, all the three parties can retrieve the information sent by David, that too in an error-immune manner. Note that the three participants can retrieve information only if all the three collaborate. Thus, the present protocol provides an illustration of the transfer of error-free information even through atmospheric turbulence with OAM modes of light.

The generalisation of this protocol to a higher number of parties is straightforward.

## IV. Conclusion

We have presented a protocol that allows to realise collaborative error-immune information transfer with OAM modes of light through weak atmospheric turbulence. Since `three qubit and qutrit' OAM states have already been realised experimentally [9], the proposed protocol can be realised with current technology.